\documentclass[fleqn,10pt]{wlscirep}
\usepackage[utf8]{inputenc}
\usepackage[T1]{fontenc}
\usepackage{hyperref}
\usepackage[final]{pdfpages}
\makeatletter
\AtBeginDocument{\let\LS@rot\@undefined}
\makeatother
\hypersetup{
    colorlinks = false,
    linkbordercolor = {white},
}

\title{Evolving spatiotemporal patterns and urban scaling of deaths from external causes}
\author[1,2]{Cesar I. N. Sampaio Filho}%cesar@fisica.ufc.br 
\author[1,2]{Humberto A. Carmona}%carmona@fisica.ufc.br
\author[3,4]{Antonio S. Lima Neto}%tanta26@yahoo.com
\author[5]{Monica~V.~Prates}%moniprates22@gmail.com
\author[5,*]{Haroldo~V.~Ribeiro}%hvribeiro@uem.br
\author[6]{Marcia C. Castro}%mcastro@hsph.harvard.edu
\author[1,2,*]{Jos\'e S. Andrade Jr.}%soares@fisica.ufc.br

\affil[1]{\small Departamento de F\'isica, Universidade Federal do Cear\'a, 60451-970 Fortaleza, Ceará, Brazil}
\affil[2]{\small Escola de Sa\'ude P\'ublica do Cear\'a, 60165-090, Fortaleza, Cear\'a, Brazil}
\affil[3]{\small Laborat\'orio de Ci\^encia de Dados e Intelig\^encia Artificial, Universidade de Fortaleza, Fortaleza, Cear\'a, Brazil}
\affil[4]{\small Secretaria Executiva de Vigil\^ancia em Sa\'ude, Secretaria da Sa\'ude do Cear\'a, Fortaleza, Cear\'a, Brazil}
\affil[5]{\small Departamento de F\'isica, Universidade Estadual de Maring\'a, Maring\'a PR 87020-900, Brazil}
\affil[6]{\small Department of Global Health and Population, Harvard T. H. Chan School of Public Health, Boston, Massachusetts, USA}

\affil[*]{\small email: hvribeiro@uem.br; soares@fisica.ufc.br}
\begin{abstract}
Urban scaling theory posits that urban indicators follow power‐law relations with population, yet the evolution of these patterns -- and the role of regional differences in settings marked by social inequalities and unplanned urbanization -- remains poorly understood. Here, we analyze nearly three decades of mortality data from Brazilian cities to investigate the scaling of external causes of death: homicides, suicides, and accidents. Using a hierarchical Bayesian framework and spatial correlation analysis, we find that these mortality indicators exhibit distinct, regionally heterogeneous scaling trajectories. Homicide mortality has significantly attenuated its typical superlinear scaling with increased spatial clustering, suggesting a redistribution of violence to smaller cities and intensified intercity interactions, possibly linked to the consolidation of organized crime. Suicide mortality, usually sublinear, has trended upward, implying a weakening of urban agglomerations' protective effect. Accident mortality remains superlinear, with transport fatalities scaling nearly proportionally, and non-transport accidents becoming superlinear. The scaling changes for suicides and accidents coincide with less correlated and stable spatial patterns, suggesting that the underlying processes predominantly operate within city boundaries. Finally, while scaling exponents have evolved more homogeneously across Brazilian states, scale-adjusted mortality remains highly heterogeneous, indicating that fundamental processes govern scaling laws, whereas state‐specific factors drive scale‐adjusted metrics.
\end{abstract}

\begin{document}
\rfoot{\small\sffamily\bfseries\thepage/16}%

\flushbottom
\maketitle
\thispagestyle{empty}

\section*{Introduction}

Urban scaling has emerged as a fundamental paradigm for understanding how city size influences a diverse array of urban phenomena~\cite{kuhnert2006scaling, bettencourt2007growth, bettencourt2013origins, west2017scale, dacci2025urban}, revealing both universal patterns and deviations across different urban systems. Drawing inspiration from biological allometry~\cite{west1997general, enquist1998allometric} -- as exemplified by Kleiber's $3/4$ law~\cite{kleiber1932body} -- and numerous scaling relationships in complex systems~\cite{thurner2018scaling}, this framework asserts that urban indicators (city properties) $Y$ scale with population size $N$ according to a power-law relation, $Y\sim N^\beta$, characterized by the urban scaling exponent $\beta$~\cite{bettencourt2007growth}. Such nonlinear scaling encapsulates the intensification of social interactions and growing complexity intrinsic to urban expansion, which consequently drive agglomeration effects that amplify productivity, innovation, and wealth creation ($\beta>1$), while simultaneously fostering economies of scale in urban infrastructure indicators ($\beta<1$).

Not all urban metrics, however, adhere to favorable scaling patterns. Indicators related to social inequality, unemployment, crime, disease transmission, and environmental degradation commonly exhibit superlinear scaling ($\beta>1$) with population size~\cite{bettencourt2010urban, alves2013distance, oliveira2014large, patterson2015per, rocha2015non, meirelles2018evolution, la2024urban}, exposing vulnerabilities that may disproportionately affect large urban areas~\cite{dye2008health}. Among these adverse urban indicators, external causes of death -- fatalities resulting from events or circumstances unrelated to internal health conditions or natural diseases -- represent a crucial category of indicators for understanding the uneven distribution of costs and risks associated with urban growth. This category includes deaths from homicides (intentional deaths caused by another person), suicides (intentional self-inflicted deaths), and accidents (unintentional fatal injuries), each reflecting complex interactions between behavioral, socioeconomic, and urbanization-related factors.  

While previous studies have used the urban scaling framework to investigate how population size affects external causes of death~\cite{bettencourt2010urban, gomez2012statistics, alves2013distance, melo2014statistical, bilal2021scaling, mcculley2022urban}, the existence of potential temporal patterns reflecting the dynamic nature of society and the role of regional differences within an urban system remain largely unexplored. This gap is especially pressing in developing countries such as Brazil, where stark social inequalities, rapid and unplanned urbanization, and regional disparities in policy implementation intersect to create a highly complex and evolving context. Here, we integrate urban scaling with spatial correlation analysis to bridge this gap, investigating mortality from homicides, suicides, and accidents across all Brazilian cities over a 28-year period (1996-2023). This interval encompasses profound demographic transitions, shifts in governmental and public policy, the COVID-19 pandemic, and substantial transformations in the urban landscapes, including disorderly and regionally heterogeneous urban expansion~\cite{mapbiomas2023cities}, growth of informal settlements~\cite{mapbiomas2022brazil}, the consolidation of organized crime networks~\cite{stahlberg2022prison}, and the dissemination of violent crime into smaller urban areas~\cite{steeves2015interiorization}. Using a hierarchical Bayesian framework, our approach accounts for state-level regional differences, enabling the identification of overarching temporal trends and region-specific nuances. Beyond urban scaling, spatial correlation analyses uncover spatial dependencies and interactions influencing mortality patterns, enriching our understanding of the dynamics underlying observed scaling trends.

Our findings reveal marked shifts in mortality rates, spatial patterns, and scaling behaviors over the past three decades. Suicide rates consistently increased, while homicide and accident mortality displayed more complex trends, including a recent decline in homicide rates and a rise in accident mortality. The superlinear scaling exponent of homicides, indicative of disproportionately high homicide rates in larger urban areas -- often viewed as a hallmark of urban violence~\cite{gomez2012statistics, alves2013distance} -- initially increased over the first two decades but has since declined, becoming less superlinear in recent years. Meanwhile, suicide mortality, traditionally displaying sublinear scaling attributed to protective effects of urban agglomeration~\cite{melo2014statistical}, has seen its exponent steadily increase, thus becoming less sublinear over time. Accident mortality exhibited stable superlinear scaling overall but revealed notable divergences when separated into transport-related and non-transport categories; the former scaled approximately linearly, while the latter scaled superlinearly, with increasing exponents in recent years. Spatial correlations in homicide rates strengthened markedly over time, contrasting with the comparatively stable and weaker spatial dependencies observed for suicide and accident mortality, reflecting distinct underlying urban processes. Although these scaling patterns were broadly consistent across Brazilian regions, significant regional disparities in scale-adjusted mortality emerged, emphasizing diverse local trajectories and uneven burdens of external-cause mortality within the urban system.

\section*{Results}

\begin{figure*}[!ht]
\includegraphics*[width=0.74\textwidth]{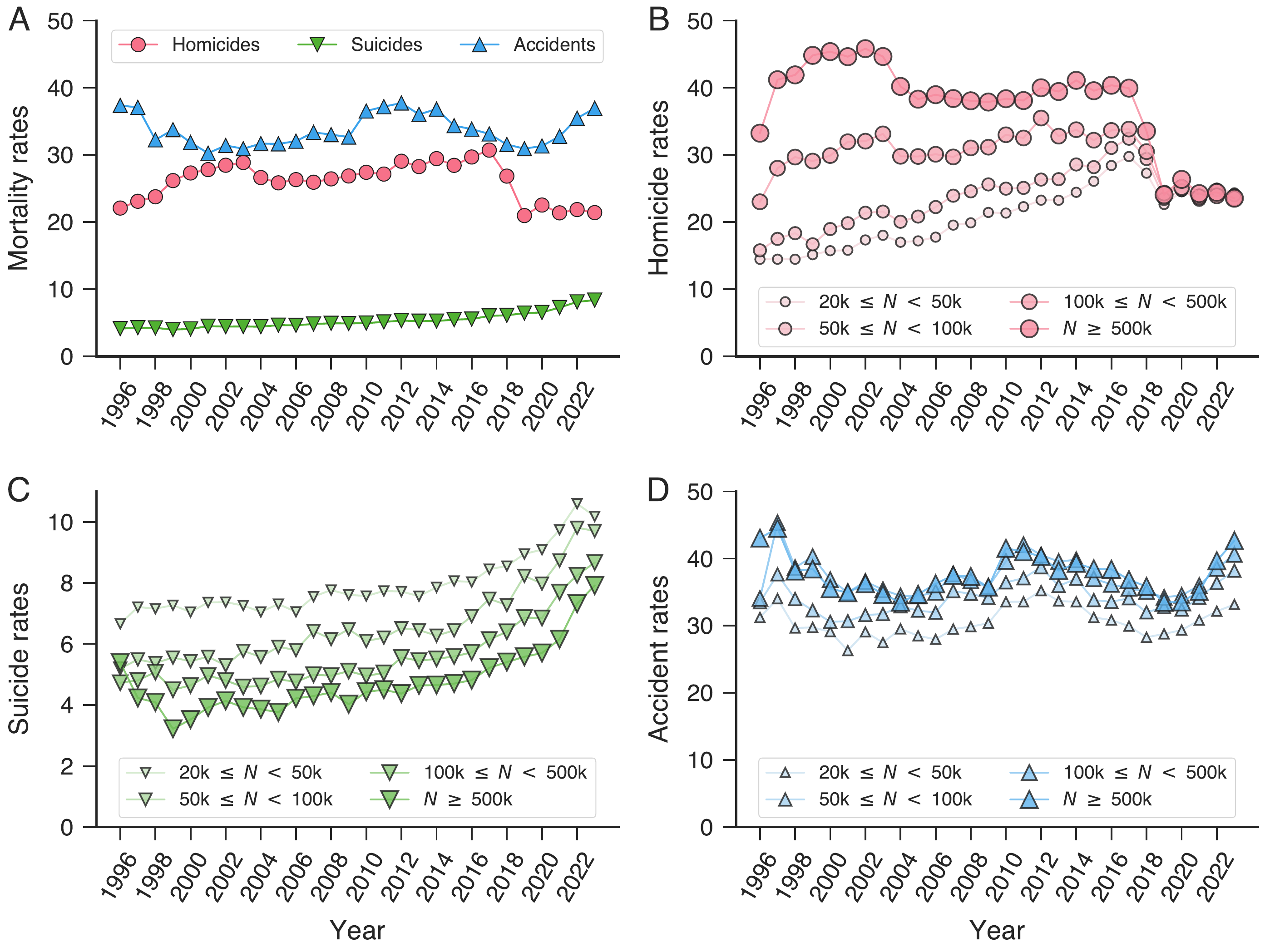}
\centering
\caption{Evolution of mortality rates from external causes in Brazil between 1996 and 2023. (A) Nationwide mortality rates due to homicides (red circles), suicides (green downward triangles), and accidents (blue upward triangles). Homicide rates showed an increasing trend from 1996 to 2006, followed by a temporary dip in 2004 and 2005 that started a prolonged period of increase until 2017, after which rates declined sharply -- from $31$ to $21$ deaths per $100{,}000$ individuals between 2018 and 2019 -- stabilizing around this value ever since. Suicide rates exhibited a monotonically increasing trend, rising from $4.1$ to $8.4$ deaths per $100{,}000$ people over the study period. Accident rates initially decreased to a minimum of $30.3$ deaths per $100{,}000$ people in 2001, followed by an increasing trend that resulted in a peak of $37.7$ deaths per $100{,}000$ people in 2012, after which rates decreased to $31$ deaths per $100{,}000$ people in 2019, initiating a new upward trend. Panels (B), (C), and (D) present mortality rates for each external cause of death stratified by city population size, grouped into four categories as indicated in the legends. (B) Homicide rates exhibited a size-dependent hierarchy until 2017, with larger cities consistently having higher rates. After 2017, rates across all size categories converged, suggesting city size ceased to be a determinant of homicide mortality. Before this convergence, smaller cities displayed a more pronounced rise, peaking in 2017, while larger cities showed a steeper decline thereafter. (C) Suicide rates maintained a stable, size-dependent hierarchy, with smaller cities experiencing consistently higher rates than larger ones. All city size categories displayed a similar increasing trend, mirroring the national pattern, although larger cities experienced a faster rise in suicide rates, particularly after 2016. (D) Accident rates vary less across city size categories, with smaller cities typically experiencing slightly lower rates and an evolution closely following the overall national trend.}
\label{fig:1}
\end{figure*}

We begin by analyzing the nationwide evolution of mortality rates (number of deaths per $100{,}000$ inhabitants) due to homicides, suicides, and accidents over the 28 years encompassed by our dataset (see \hyperref[sec:methods_data]{Methods} for details). Figure~\ref{fig:1}A depicts these rates as a function of time, revealing distinct trajectories for each cause. Homicide rates exhibited two successive upward trends, separated by a transient dip in 2004-2005. The rate peaked at $29$ deaths per $100{,}000$ individuals in 2003 and reached a more recent maximum of $31$ deaths per $100{,}000$ individuals in 2017. This was followed by a substantial decline over the subsequent two years, after which rates stabilized at approximately $22$ deaths per $100{,}000$ individuals post‐2019  -- a value markedly higher than the global average of roughly $5$ deaths per $100{,}000$ people as of 2023~\cite{unodc2025homicide}. In contrast, suicide rates increased steadily, rising from $4.1$ deaths per $100{,}000$ individuals in 2009 to $8.4$ in 2023, thereby approaching the global rate of $9.2$ as of 2020~\cite{who2025suicide}. Mortality from accidents followed a more complex trajectory, initially declining to a minimum of $30.3$ deaths per $100{,}000$ individuals in 2001, rising to $37.7$ in 2012, and then decreasing again to $31$ in 2019, which marked the onset of a renewed upward trend. Despite the convergence between homicide and accident mortality rates in the early 2000s and again in the late 2010s, the overall ranking of external causes of death remained stable, with accident-related mortality consistently the highest, followed by homicides, and suicide rates remaining significantly lower throughout the study period.

We also investigate the evolution of mortality rates by grouping Brazilian cities into four population size ($N$) categories: small ($20{,}000 \leq N < 50{,}000$), intermediate ($50{,}000 \leq N < 100{,}000$), large ($100{,}000 \leq N < 500{,}000$), and major ($N \geq 500{,}000$) urban areas. Figures~\ref{fig:1}B-D present the estimated mortality rates over time for each external cause of death across these categories, highlighting significant differences in both aggregate trends and stratified patterns by city size. Before 2017, homicide rates exhibited a clear hierarchy, with larger cities consistently reporting higher rates. During this period, small and intermediate urban areas experienced sharper increases in homicide rates compared to large and major urban areas. Following 2017, aligned with national trends, homicide rates declined across all city categories, with the decrease being more pronounced in larger cities. By 2018, this decline resulted in a convergence of homicide rates, which became nearly indistinguishable across the different city-size groups. Taken together, these findings suggest a spread of homicides to small and intermediate cities between 1994 and 2017, with major cities showing comparatively stable levels. Thereafter, the decline -- stronger among larger cities -- and the convergence of homicide rates across all city sizes indicate a system-wide reduction of this type of violence, particularly among larger cities and metropolitan areas. Suicide rates followed a pattern similar to that observed nationally, maintaining a clear hierarchy in which smaller cities consistently exhibited higher rates throughout the study period, although this disparity diminished owing to a faster rise in rates in larger cities, particularly after 2016. Moreover, all city-sized groups presented a faster growth in suicide rates after 2020, a pattern that is likely associated with the mental-health impacts of the COVID-19 pandemic~\cite{melo2025impact}. Accident rates evolved in close alignment with national trends but showed less variation among the city-size categories, particularly between large and major urban areas, while small and intermediate cities typically registered lower rates.

\begin{figure*}[!ht]
\includegraphics*[width=1\textwidth]{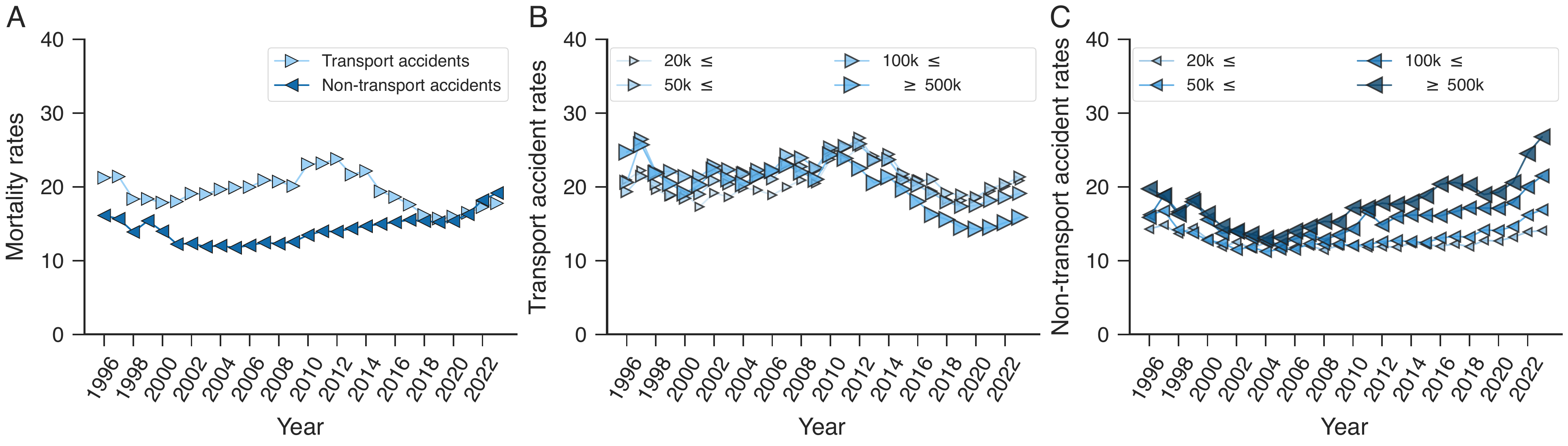}
\centering
\caption{Evolution of mortality rates from transport and non‐transport accidents in Brazil between 1996 and 2023. (A) Nationwide mortality rates for transport (light blue rightward triangles) and non‐transport (dark blue leftward triangles) accidents. Transport-related mortality exceeded that from non‐transport accidents until 2018, with a more pronounced gap during the 2000s and early 2010s. Transport accident rates began declining in 2012, whereas non‐transport rates exhibited a small but steady upward trend, leading to convergence in 2018. After that year, both rates followed a similarly slight upward trajectory. Panels (B) and (C) present mortality rates for transport and non‐transport accidents stratified by city population size, grouped into four categories as indicated in the legends. (B) Transport accident rates were less dependent on city size and followed trends similar to those observed nationwide. (C) Non‐transport accident rates showed a clear dependence on city size, with larger cities exhibiting significantly higher rates, particularly after 2010.}
\label{fig:2}
\end{figure*}

To further elucidate the dynamics of accident-related mortality, we disaggregate accident rates by distinguishing between transport and non-transport accidents (see \hyperref[sec:methods_data]{Methods} for details). Figure~\ref{fig:2}A shows the nationwide evolution of both mortality rates, while Figures~\ref{fig:2}B-C depict the stratified patterns by population according to the same city‐size categories used in our previous analysis. Transport accidents exhibited higher rates than non‐transport accidents until 2018. This gap was particularly pronounced during the 2000s and early 2010s when transport accidents displayed a steeper increasing trend. After 2012, however, transport accident mortality began to decline, falling from $23.8$ deaths per $100{,}000$ people to $15.7$ in 2019, whereas non-transport accident mortality maintained a slightly increasing trend. This decrease in transport accident mortality likely reflects the impact of Brazil's \textit{``Lei Seca''} (literally, ``Dry Law''), a legislative measure introduced in June 2008 that prohibits alcohol consumption by drivers and enforces a strict zero-tolerance policy for blood alcohol concentration. The gradual enhancement of enforcement measures and the consequent behavioral adaptations among drivers may explain the observed lag between the law's implementation and the decline in transport-related mortality. Consequently, the rates converged after 2018, each exhibiting a similar upward trajectory in recent years. The city‐size stratified patterns mirrored the nationwide trends, although transport accident rates were less dependent on city size; rates in major urban areas became significantly lower than those in other categories only in recent years. Conversely, non‐transport accident rates exhibited a clearer hierarchy, with larger cities consistently reporting higher rates, particularly after 2010.

The stratified patterns in homicide and suicide mortality rates across city sizes -- and the absence of a clear distinction in transport accident rates among these groups -- are consistent with previous findings that homicides scale superlinearly~\cite{alves2013distance}, suicides sublinearly~\cite{melo2014statistical}, and transport accidents approximately linearly~\cite{melo2014statistical} with population size. However, previous studies have not explicitly accounted for regional disparities, considered the distinction between transport and non-transport accidents, or examined whether the scaling regimes of external mortality causes have remained static over time. Addressing these gaps is particularly important in developing and socioeconomically heterogeneous countries, such as Brazil, and is essential for identifying potential shifts in scaling regimes. The convergence of homicide rates across city-size categories observed after 2018, the more pronounced upward trend in suicide rates among larger cities in recent years, and the marked discrepancies in non-transport accident rates among city-size categories after 2010 suggest such shifts. 

To investigate these issues, we analyze the urban scaling of external causes of death using a Bayesian hierarchical approach, in which the population ($N$) and the number of deaths ($Y$) for each city are nested within Brazilian states. Specifically, we model
\begin{equation}\label{eq:bayesian_scaling}
  \log Y_j \sim \mathcal{N}(a_j + \beta_{j} \log N_j, \varepsilon)\,,
\end{equation}
where $j$ indexes each Brazilian state, $\mathcal{N}(\mu,\sigma)$ denotes a normal distribution with mean $\mu$ and standard deviation $\sigma$, and $\varepsilon$ represents the unexplained variance in $\log Y_j$. We further assume
$\beta_j \sim \mathcal{N}(\beta, \sigma_\beta)$ and $a_j \sim \mathcal{N}( a, \sigma_a)$, where $\beta$ and $a$ are the means, and $\sigma_\beta$ and $\sigma_a$ are the standard deviations of the normal distributions associated with urban scaling exponents $\beta_j$ and prefactors $a_j$, respectively. The Bayesian inference process consists of estimating the posterior probability distributions of the parameters at both the national level ($\beta$, $ a$, $\sigma_\beta$, and $\sigma_a$) and the state level ($\beta_j$ and $a_j$ for each state $j$). In simplified terms, for a given year, the point estimate for $\beta$ corresponds to the urban scaling exponent for the entire country, while the point estimates for $\beta_j$ represent the urban scaling exponents for individual states. Additionally, Brazilian cities are defined as administrative units (municipalities) and, as such, may not all conform to the concept of functional urban areas, particularly small towns. Smaller settlements may lack the socioeconomic complexity and agglomeration effects typical of larger urban areas, potentially leading to inaccuracies in our scaling analyses. We address this issue by following other urban scaling studies~\cite{alves2014empirical, ribeiro2021association} and considering only cities that belong to the power-law regime of the population distribution, that is, cities adhering to Zipf's law~\cite{zipfhuman2012, auerbacghesetz1913}. To do so, we apply the Clauset-Shalizi-Newman method~\cite{clauset2009power, alstott2014powerlaw}, which simultaneously estimates the Zipf's law exponent $\alpha$ and lower bound $N^*$ of the power-law regime of the population distribution. We then use $N^*$ as the population threshold for selecting cities eligible for our scaling results. As shown in Supplementary Figure~S1, the values of $N^*$ display slight fluctuations around $20{,}000$, while Zipf's law exponents are very close to $1.1$ over the years. Thus, for each year, we adjust the model of Equation~\ref{eq:bayesian_scaling}, considering cities with $N\geq N^*$ (see \hyperref[sec:methods_model]{Methods} for details). To ensure our results are not overly dependent on this specific cutoff, we perform a sensitivity analysis using alternative population thresholds, which confirms that the estimated scaling exponents are robust to this choice (Supplementary Figure~S2).

\begin{figure*}[!ht]
\includegraphics*[width=1\textwidth]{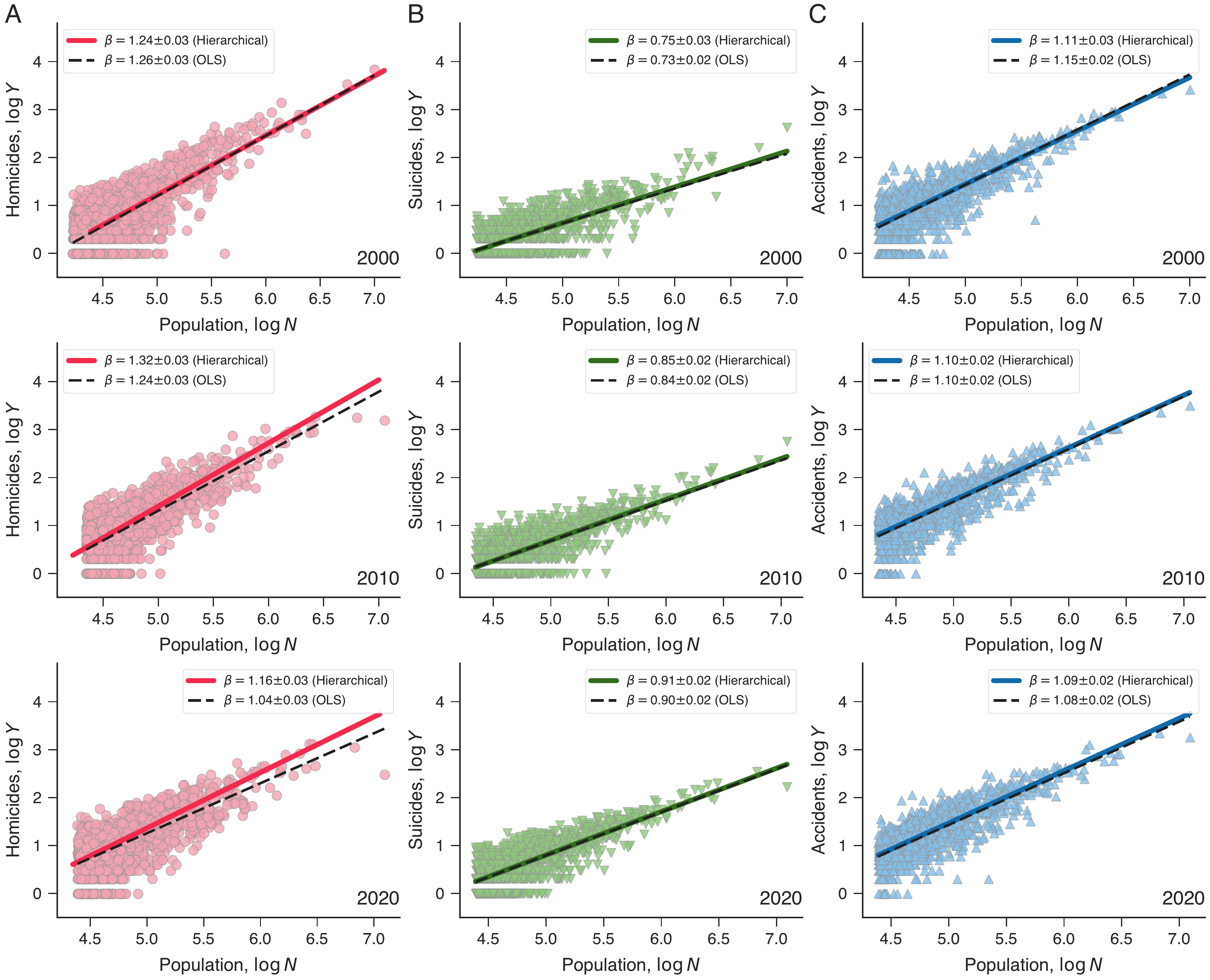}
\centering
\caption{Urban scaling of external causes of death in Brazil. Population scaling relations for (A) homicides, (B) suicides, and (C) accidents in 2000, 2010, and 2020. Markers represent the number of deaths ($Y$) due to each external cause versus the population ($N$) of Brazilian cities on a base-10 logarithmic scale ($\log Y$ versus $\log N$). Continuous lines indicate the nationwide scaling laws estimated using our Bayesian hierarchical approach, while dashed lines represent scaling laws obtained via the ordinary least-squares (OLS) method applied to log-transformed data. The urban scaling exponents and their standard errors are provided in the plot legends. The Bayesian estimated scaling laws closely match the OLS estimates for suicides and accidents, whereas for homicides, the Bayesian approach consistently yields a higher scaling exponent across all years. These findings indicate that homicides scale superlinearly, suicides sublinearly, and accidents linearly with population size, with the scaling exponent for homicides exhibiting a significant decline over time.}
\label{fig:3}
\end{figure*}

Figure~\ref{fig:3} illustrates examples of scaling relations for homicides, suicides, and accidents in three different years (2000, 2010, and 2020), along with the scaling laws derived for the entire country (continuous lines) compared with standard ordinary least-squares estimates (dashed lines). Supplementary Figures~S3-S7 present the results for all years -- including those obtained after disaggregating accidents into transport‐related and non‐transport‐related categories. The Bayesian approach yields scaling laws similar to those obtained using the ordinary least-squares method for suicides and accidents (the same holds for transport‐related and non‐transport‐related accidents), whereas it consistently produces a higher scaling exponent for homicides. This finding indicates that regional differences are more pronounced for homicides than for other external causes of death. Given our interest in these regional differences, we focus our analysis on the scaling laws estimated using the Bayesian approach. Figure~\ref{fig:3} further indicates that the scaling laws have significantly changed for homicides and suicides, with nationwide scaling exponents for homicides becoming less superlinear and the exponent for suicides becoming less sublinear, whereas the exponents for accidents have remained nearly constant over these three years. Moreover, the quality of the scaling relations has remained stable over time, with Pearson and Spearman correlation coefficients (between $\log Y$ and $\log N$) oscillating around $0.8$ and $0.7$, respectively (Supplementary Figure~S8). This stability indicates that the observed changes are not attributable to an improvement or deterioration of the scaling relations.

\begin{figure*}[!ht]
\includegraphics*[width=0.9\textwidth]{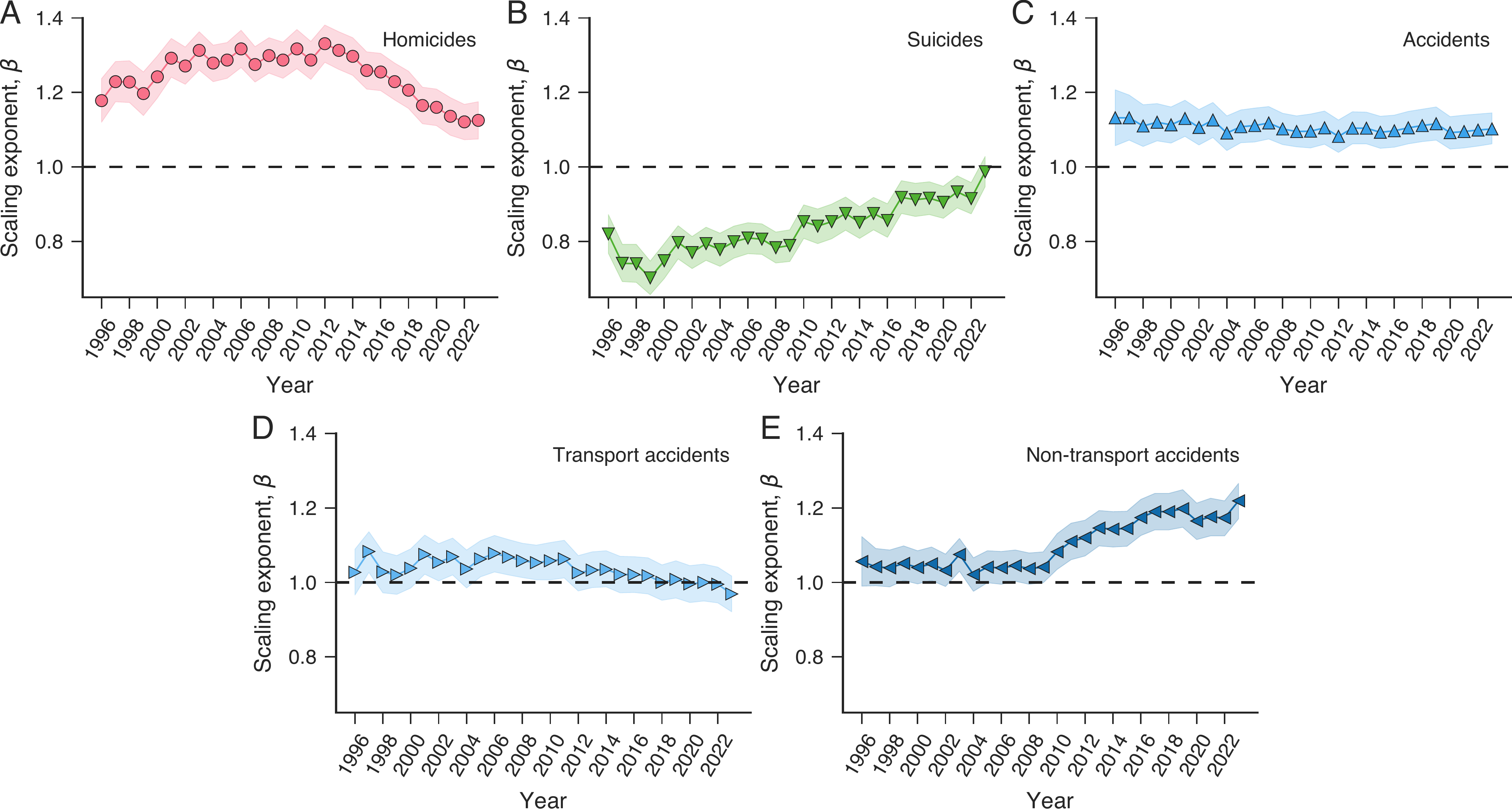}
\centering
\caption{Evolution of the urban scaling exponent for external causes of death in Brazil between 1996 and 2023. (A) Homicide exponents initially increased from $\beta=1.17$ in 1996 to a peak of $\beta=1.33$ in 2012, then decreased sharply to a minimum of $\beta=1.12$ in 2022. (B) Suicide exponents exhibited an overall upward trend, rising from a minimum of $\beta=0.70$ in 1999 to $\beta=0.92$ in 2022. (C) Accident exponents remained relatively stable over the years, oscillating around $\beta\approx1.1$. Disaggregating accidents into (D) transport-related and (E) non-transport-related categories reveals distinct trends. Transport-related accidents display an almost constant exponent near $1$, whereas the exponents for non-transport-related accidents, which were close to $1$ until 2010, increased thereafter, reaching a plateau at approximately $\beta\approx1.2$ in recent years. In all panels, markers denote the mean of the posterior distribution of $\beta$, and the shaded regions represent the 96\% highest density (credible) intervals, with dashed lines indicating the isometric scaling regime.}
\label{fig:4}
\end{figure*}

We investigate the nationwide evolution of scaling laws using the point estimates of the scaling exponents $\beta$ for the three external causes of death, in addition to disaggregating accident-related fatalities into transport and non-transport categories. Figure~\ref{fig:4} shows that each mortality type exhibits distinct temporal trends in its scaling exponents. Homicide scaling exponents, which remained superlinear throughout the study period (Figure~\ref{fig:4}A), increased from $\beta=1.17$ in 1996 to $\beta=1.33$ in 2012, then decreased sharply over the subsequent decade to a minimum of $\beta=1.12$ in 2022. This recent decrease in the scaling exponent indicates that large cities have become proportionally less violent. For instance, a $1$\% increase in city population was associated with a $1.33$\% increase in the number of murders in 2012, compared with a $1.12$\% increase in 2022. To illustrate the magnitude of these changes, an increase in population from 100,000 to 500,000 inhabitants corresponded to an average rise from approximately $27$ to $230$ homicide cases in 2012, whereas the same change corresponded to an average increase from approximately $22$ to $134$ homicide cases in 2022. In contrast, suicide exponents display an increasing trend starting from a sublinear baseline (Figure~\ref{fig:4}B), with values rising almost monotonically from a minimum of $\beta=0.70$ in 1999 to $\beta=0.92$ in 2022. This shift represents a marked reduction in the ``protective effect'' of larger urban agglomerations against self‐directed violence, an effect that has been attributed to the proportionally larger social connections available in bigger cities~\cite{schlapfer2014scaling, melo2014statistical}. In 1999, a $1$\% rise in population was associated with only $0.70$\% rise in the average number of suicides, whereas that same change in 2022 corresponded to a $0.92$\% increase. Using the same illustrative example, moving from a city of 100,000 inhabitants to one with 500,000 was associated with an average increase from approximately $4$ to $13$ suicide cases in 1999 and from $8$ to $35$ cases in 2022.

Accident scaling exponents remained stable in a superlinear regime, with $\beta\approx1.1$ over the 28-year period covered by our dataset (Figure~\ref{fig:4}C). Consequently, residents of larger cities face proportionally higher risks of fatal accidents. On average, a 1\% increase in population was associated with a 1.1\% rise in accident-related deaths throughout the study period. In 2022, the superlinear scaling implied that transitioning from a city with 100,000 inhabitants to one with 500,000 would correspond to an increase in fatalities from approximately $31$ to $184$, compared with an increase from $31$ to $157$ deaths expected under the isometric regime. However, when accidents are disaggregated into transport and non‐transport categories, distinct scaling behaviors emerge. The scaling exponents for transport‐related accidents (Figure~\ref{fig:4}D) are closer to unity than those obtained from the aggregated data, particularly after 2012, when estimates of $\beta$ are statistically indistinguishable from 1. Thus, in agreement with previous studies~\cite{melo2014statistical}, our findings indicate that transport accidents scale approximately isometrically with population and that this linear regime remained stable over time. In contrast, non‐transport‐related accidents (Figure~\ref{fig:4}E), which exhibited scaling exponents near $1$ between 1996 and 2010, increased significantly thereafter and began to exhibit a superlinear regime with $\beta\approx1.2$ in more recent years -- a change that drastically affects the expected raw number of deaths. For example, a population increase from 100,000 to 500,000 was associated with an increase from approximately 11 to 58 deaths related to non‐transport accidents in 2002 and from 14 to 90 in 2022.

\begin{figure*}[!ht]
\includegraphics*[width=0.85\textwidth]{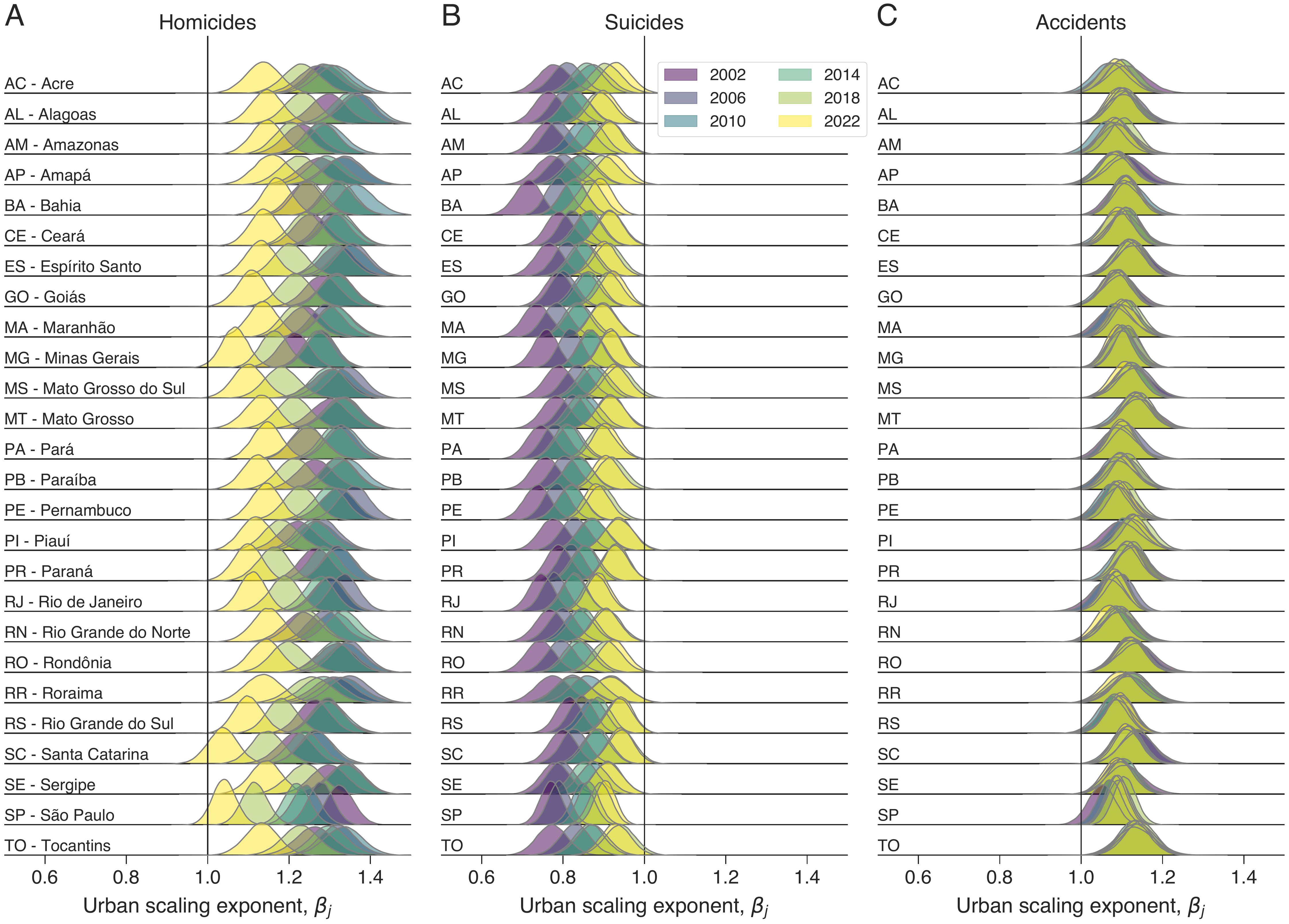}
\centering
\caption{Individual evolution of mortality scaling exponents for each Brazilian state from 1996 to 2023. Colored curves represent the posterior probability distribution of $\beta_j$ for each state $j$, estimated for (A) homicides, (B) suicides, and (C) accidents. Each row of plots corresponds to a Brazilian state, with colors indicating different years (as shown in the legend). Vertical lines in all three panels denote the isometric regime. The evolution of each state generally aligns with nationwide trends for homicides and suicides, although some states exhibit distinct scaling exponents and rates of change. Consistent with national patterns, the posterior distributions of $\beta_j$ remain approximately stationary and show little variation across states.}  
\label{fig:5}
\end{figure*}

Having characterized nationwide trends, we now focus on the evolution of mortality scaling exponents at the state level. Figure~\ref{fig:5} shows the posterior probability distributions of state‐specific exponents $\beta_j$ for homicides, suicides, and accidents (Supplementary Figure~S9 further disaggregates accidents into transport and non-transport fatalities). Each row depicts the distributions for a Brazilian state (see Supplementary Figure~S10 for their geographic location), and colors indicate six evenly spaced years (see Supplementary Figures~S11 and~S12 for the point estimates of $\beta_j$ across all 28 years). The evolution of these distributions reinforces the robustness of the nationwide trends, with the trajectories of Brazilian states following the country's overall behavior. For homicides, most exponent distributions lie entirely in the superlinear region, shifting toward values closer to unity in recent years (Figure~\ref{fig:5}A). In contrast, the exponent distributions for suicides have increased over time while remaining predominantly in the sublinear regime (Figure~\ref{fig:5}B). The exponent distributions for accidents consistently exhibit stable superlinear behavior (Figure~\ref{fig:5}C); further disaggregation of accident‐related fatalities into transport and non-transport categories reveals that the exponent distributions for transport accidents cluster around $\beta_j=1$ (Supplementary Figure~S9A), whereas those for non‐transport accidents have shifted into the superlinear region over time (Supplementary Figure~S9B).

However, inter‐state variability exists both in the exponents and in the pace of change, particularly for homicides. For example, states such as S\~ao Paulo (SP), Santa Catarina (SC), and Minas Gerais (MG) have closely approached the isometric regime, whereas states such as Bahia (BA) and Pernambuco (PE) have retained higher superlinear exponents. We quantified the disparities among states' exponent distributions by calculating the percentage of state pairs with non-overlapping interquartile ranges. The results (Supplementary Figure~S13) confirm greater inter‐state variability for homicides, with approximately $25$\% of state pairs exhibiting non-overlapping interquartile ranges. In contrast, the distributions for suicides and transport accidents show non-overlapping interquartile ranges in about $12$\% of pairwise comparisons, whereas accidents and non‐transport accidents exhibit non-overlapping ranges in roughly $5$\% of comparisons. Moreover, the percentages of non-overlapping interquartile ranges were relatively higher in the late 1990s for all external causes of death, declined during the early 2000s, and stabilized after the mid‐2000s (\textit{e.g.}, homicides displayed approximately $35$\% of pairwise comparisons with non-overlapping interquartile ranges in the late 1990s), indicating a general reduction in inter-state variability in scaling exponents. At the same time -- and mainly for homicides -- a few states systematically present distributions with interquartile ranges that do not overlap with those of most other states (Supplementary Figures~S14 and~S15). This occurs, for instance, with S\~ao Paulo (SP), the wealthiest and most populous state in Brazil, whose exponent distributions overlap with those of only three other states.

\begin{figure*}[!ht]
\includegraphics*[width=0.75\textwidth]{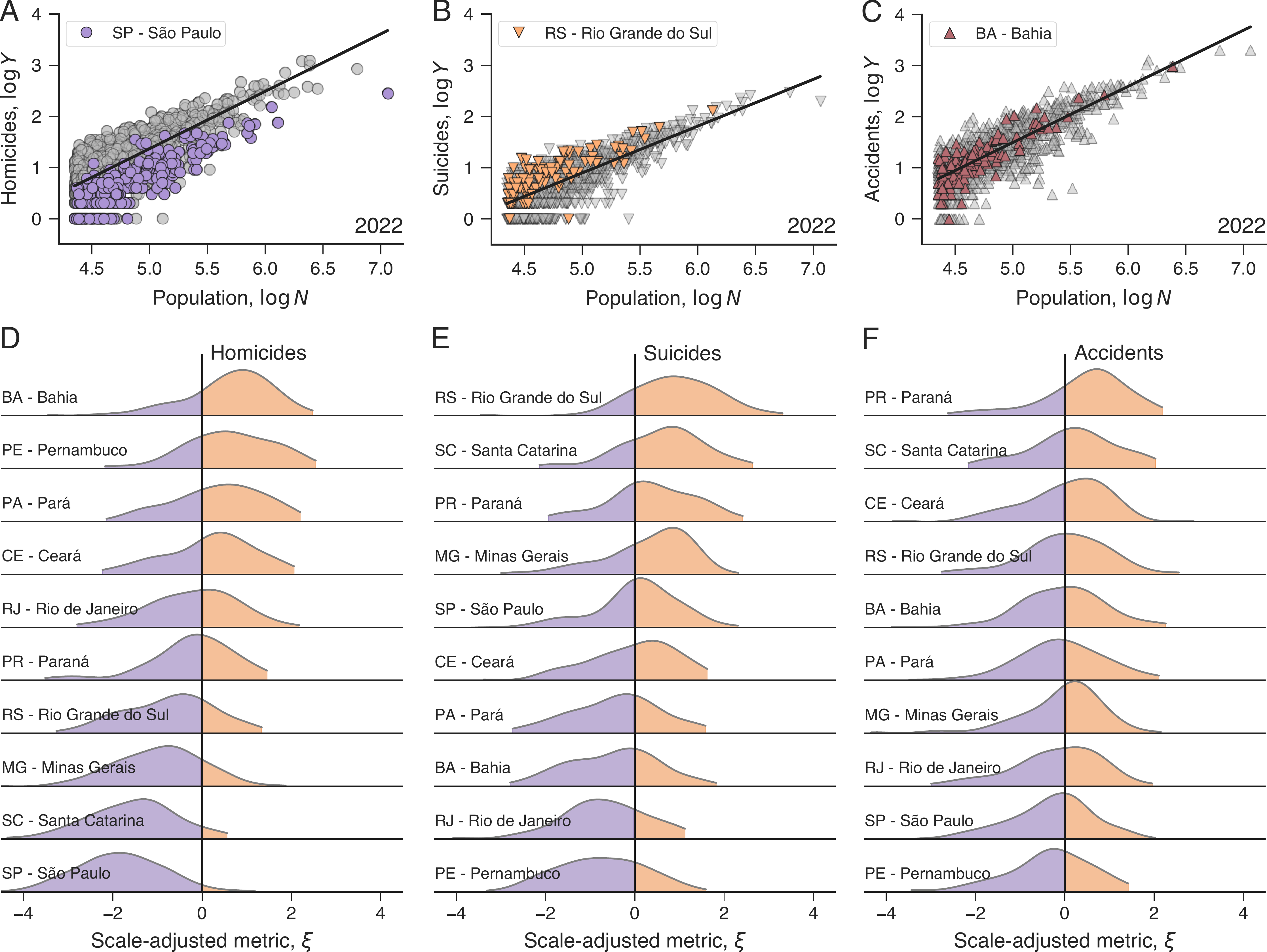}
\centering
\caption{Inter-state variability in scale-adjusted metrics. Illustrative examples of scaling relationships for (A) homicides, (B) suicides, and (C) accidents in 2022. In these panels, continuous lines represent the nationwide scaling laws estimated from our Bayesian hierarchical approach, grey markers indicate the number of deaths ($Y$) due to each external cause versus the population ($N$) of Brazilian cities on a base-10 logarithmic scale ($\log Y$ versus $\log N$), and colored markers highlight cities in the state of S\~ao Paulo (SP, circles in panel A), Rio Grande do Sul (RS, downward triangles in panel B), and Bahia (BA, upward triangles in panel C). Probability distributions of the scale-adjusted metrics $\xi$ for the ten largest Brazilian states (as indicated by their two-letter abbreviation) for (D) homicides, (E) suicides, and (F) accidents. Vertical lines separate the positive (in orange, cities exceeding scaling‑law expectations) and negative (in purple, cities falling below scaling‑law expectations) ranges of $\xi$ values.}
\label{fig:6}
\end{figure*}

In addition to inter‐state variability in the scaling exponents, we further quantify variability in the number of deaths across cities within each state relative to the expected number predicted by the country‐level scaling laws. Figure~\ref{fig:6} illustrates these differences by showing the scaling relations for homicides, suicides, and accidents in 2022, while highlighting the number of deaths reported in cities in three selected states. For homicides, cities in the state of S\~ao Paulo (SP) consistently report numbers of murders below those expected from the country‐level scaling law (Figure~\ref{fig:6}A). Conversely, cities in the state of Rio Grande do Sul (RS) present suicide cases that typically exceed the scaling law predictions (Figure~\ref{fig:6}B), whereas cities in the state of Bahia (BA) report accident fatalities distributed around the scaling law (Figure~\ref{fig:6}C). 

We quantify this other form of inter-state variability by estimating the distributions of the scale-adjusted metrics~\cite{bettencourt2010urban, lobo2013urban, alves2013distance, alves2015scale, ribeiro2024comparing} $\xi$ after grouping cities by state. This quantity is defined as the difference between the logarithm of the number of deaths in a city and the logarithm of the number of deaths predicted by the country-level scaling law, that is, $\xi = \log Y - (a+\beta \log N)$. Here $a$ and $\beta$ denote the point estimates of the prefactor and urban scaling exponent at the country level (mean of their posterior distributions), while $Y$ and $N$ correspond to the number of deaths and the population size of the city, respectively. Cities with positive values of $\xi$ present more deaths than expected from the country-level scaling laws, whereas those with negative values exhibit fewer deaths than predicted. Figures~\ref{fig:6}D-F depict the distributions of $\xi$ for homicides, suicides, and accidents in 2022 across the ten largest Brazilian states, sorted by the average value of $\xi$ for each external cause of death. Additionally, Supplementary Figure~S16 presents the temporal evolution of the average $\xi$ across all states over the 28 years covered by our dataset, while Supplementary Figure~S17 disaggregates these averages for accident fatalities into transport and non-transport accidents. 

These results reveal substantial heterogeneity in scale-adjusted mortality across states for all external causes of death, with approximately two-thirds of state pairwise comparisons exhibiting statistically significant differences in average $\xi$ (Supplementary Figure~S18). External causes of death display similar degrees of heterogeneity, with homicides exhibiting the highest fraction of significant differences ($\approx$75\%) and non-transport accidents the lowest ($\approx$60\%). Moreover, in contrast to the relatively homogeneous temporal evolution of scaling exponents across states, the evolution of average $\xi$ values is considerably more variable. In some states, the average $\xi$ evolves from negative to positive or near zero (and vice versa), whereas in others it remains largely unchanged (Supplementary Figures~S16 and~S17). The more homogeneous evolution of scaling exponents across states suggests that changes in the scaling laws result from structural and fundamental socioeconomic, infrastructural, and organizational processes operating at the national scale. Conversely, scale-adjusted metrics appear to capture state-specific factors -- including differences in local policies, cultural practices, economic conditions, and public health interventions -- which can vary considerably among Brazilian states and over time.

\begin{figure*}[!ht]
\includegraphics*[width=1\textwidth]{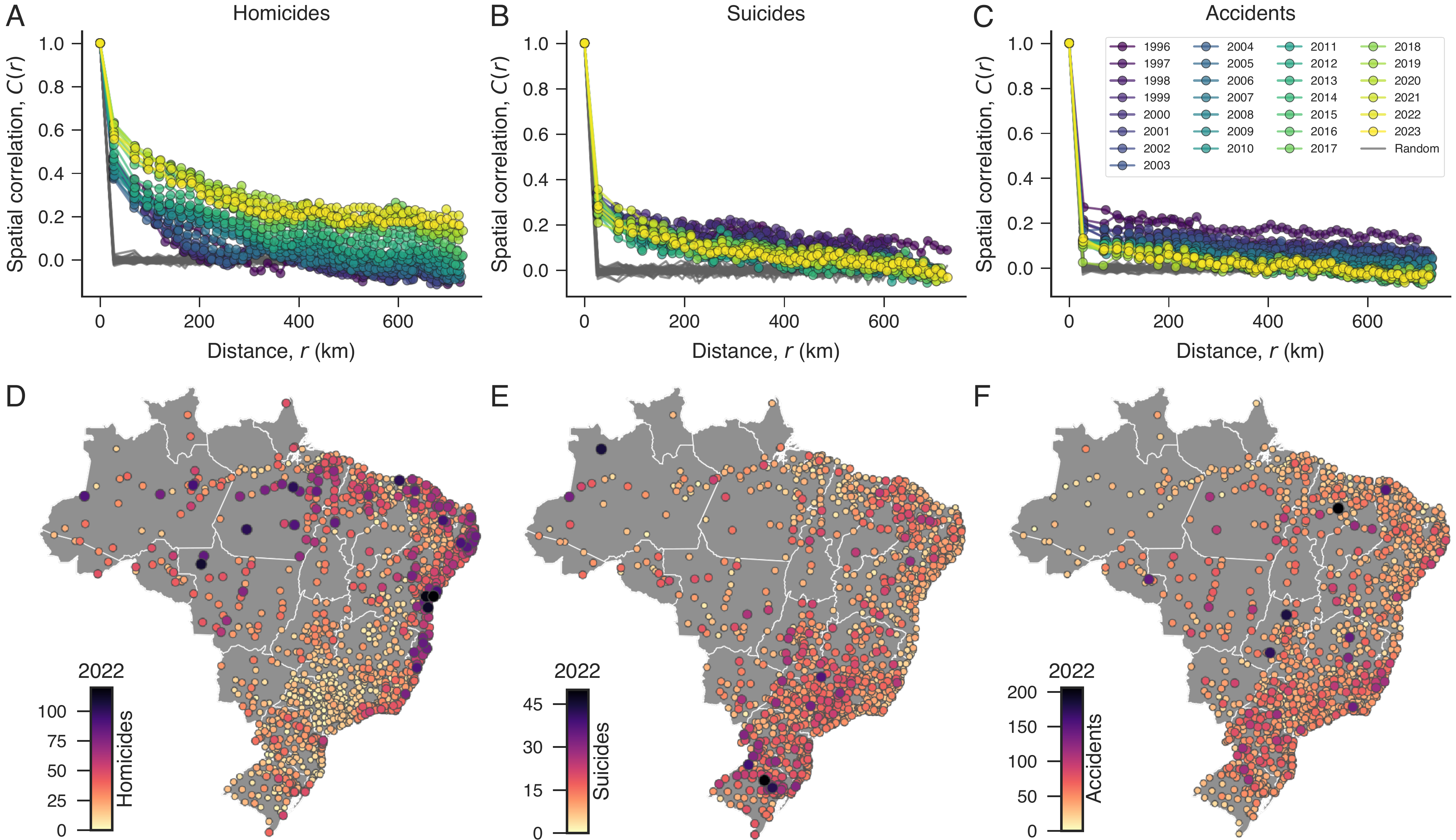}
\centering
\caption{Spatial patterns in mortality rates and their evolution. Spatial correlations $C(r)$ of mortality rates as a function of inter‐city distance $r$ for (A) homicides, (B) suicides, and (C) accidents for each year. Colored markers denote specific years (as indicated in the legend), while grey curves represent $C(r)$ values computed after randomly shuffling mortality rates among cities. The value of $C(r)$ corresponds to the Pearson correlation between the mortality rates of all city pairs whose distances fall within a range centered at $r$. Distance ranges are determined based on the percentiles of the inter‐city distance distribution, ensuring that each $C(r)$ is estimated from a comparable number of city pairs. Homicide rates exhibit markedly stronger spatial clustering; although suicide and accident correlations decline rapidly with distance, their persistence above randomized baselines also indicates nontrivial spatial organization. Illustration of the spatial distributions of mortality rates in 2022 for (D) homicides, (E) suicides, and (F) accidents. In these maps, circles indicate the geographic locations of Brazilian cities, with colors representing mortality rates and marker sizes scaled proportionally to these rates. The maps visually corroborate the correlation analysis, with homicides displaying distinct clusters, suicides an intermediate degree of clustering, and accidents a more uniform spatial distribution.} 
\label{fig:7}
\end{figure*}

To elucidate the underlying dynamics of scaling trends and their potential connection with inter-city interactions, we evaluate the spatial correlation function of mortality rates across Brazilian cities over the 28-year period covered by our data. The spatial correlation $C(r)$ quantifies the degree of association between mortality rates of cities separated by a given distance $r$, allowing us to examine spatial patterns in external causes of mortality and their temporal evolution. Conceptually, $C(r)$ is equivalent to the Pearson correlation coefficient computed for all pairs of cities approximately separated by a distance $r$ (see \hyperref[sec:methods_correlation]{Methods} for details). Positive values of $C(r)$ indicate that cities at a distance $r$ tend to exhibit similar mortality rates (for instance, cities with high rates are typically surrounded by cities with high rates), whereas negative values suggest a tendency toward dissimilarity (for instance, cities with high rates are typically adjacent to cities with low rates). Urban indicators generally exhibit positive spatial correlations that decrease as the distance between cities increases~\cite{gallos2012collective, alves2015spatial, alves2019hidden}, reflecting both a clustering tendency and the diminishing influence of one city on another as spatial separation increases. Figures~\ref{fig:7}A-C display the correlation functions for homicide, suicide, and accident rates estimated for each year, with dark bluish-purple tones corresponding to early years and bright yellow hues indicating the most recent years. In these figures, continuous gray lines denote correlation functions calculated after shuffling mortality rates among cities. 

Overall, these findings indicate that homicide rates exhibit considerably stronger spatial correlations than suicide and accident rates. Although both suicide and accident correlations decline rapidly with distance -- with the decay in suicide correlations being slower than that for accident rates -- they still decay more gradually than those observed for the shuffled data, particularly in the most recent years. Disaggregating accidents into transport and non-transport categories does not reveal pronounced differences, aside from slightly higher values for transport rates (Supplementary Figures~S19A and S19B). Figures~\ref{fig:7}D-F illustrate the spatial distributions of each external cause of death in 2022. In agreement with the correlation analysis, we observe strong clustering patterns among homicide rates, with cities exhibiting high rates typically surrounded by cities with high rates (mainly in the coastal areas of the Northeast and in the North regions), whereas cities with low rates tend to be adjacent to one another (mainly in the Southeast and particularly in the state of S\~ao Paulo). In turn, suicide rates display an intermediate degree of clustering, being more homogeneously distributed across the Brazilian territory yet featuring higher values in the Southeast and Southern regions, with significantly lower values observed elsewhere. Accident rates are most evenly distributed, with only a few small hotspots observed in the central region (Supplementary Figures~S19C and S19D indicate similar patterns for transport and non-transport accidents).

\begin{figure}[!ht]
\includegraphics*[width=0.8\textwidth]{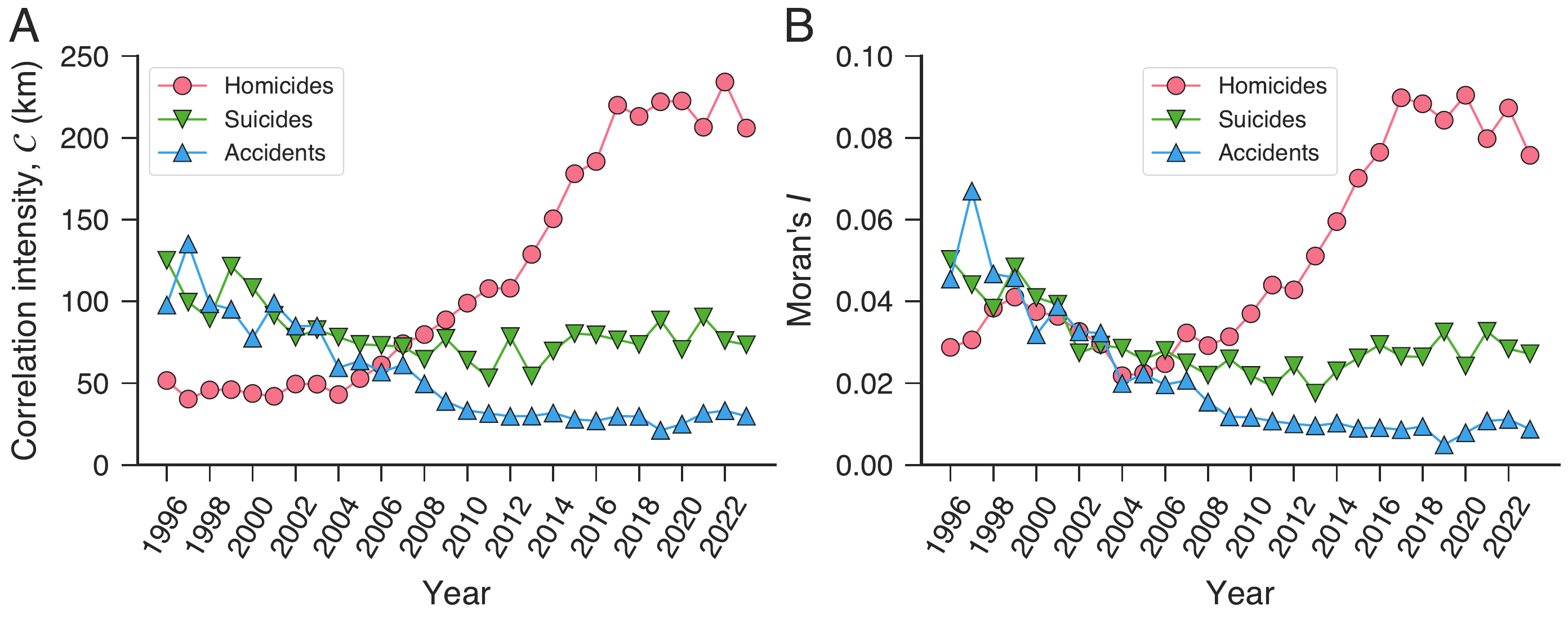}
\centering
\caption{Changes in the spatial correlations of mortality rates. (A) Evolution of spatial correlation intensity $\mathcal{C}$ and (B) Moran's $I$ spatial correlation coefficient of mortality rates across Brazilian cities for homicides (red circles), suicides (green downward triangles), and accidents (blue upward triangles). The values of $\mathcal{C}$, defined as the area under the correlation function, reveal a marked increase in spatial correlations for homicide rates, while suicides and accidents decreased and plateaued at lower levels, with Moran's $I$ corroborating these changes in spatial patterns.}  
\label{fig:8}
\end{figure}

Our correlation analysis indicates that inter-city interactions have changed over time. Specifically, $C(r)$ values for homicide rates have gradually increased, whereas in the earliest years, suicide and accident rates exhibited slightly higher correlations. To quantify these trends, we evaluate the area under the correlation function $\mathcal{C}$, providing a measure of correlation intensity across different spatial scales (larger values of $\mathcal{C}$ indicate more pronounced spatial correlations in mortality rates). Under the assumption that the correlation function follows either an exponential form [$C(r)=e^{-r/r_c}$, with $r_c$ denoting the characteristic correlation distance] or a stretched exponential form [$C(r) = e^{-r^\nu/r_c}$, where $\nu$ is the stretching exponent], the area $\mathcal{C}$ can be associated with the characteristic correlation distance ($\mathcal{C}=r_c$ in the exponential case and $\mathcal{C}\propto r_c^{1/\nu}$ in the stretched exponential case). Although these parametric functions approximate the empirical behavior of $C(r)$, we use $\mathcal{C}$ because it is directly estimated from the data without assuming a universal functional form for the correlation function across all years and mortality types, and it offers a straightforward interpretation. Figure~\ref{fig:8}A depicts the evolution of $\mathcal{C}$ for the three external causes of death. Before 2004, homicide rates exhibited the lowest correlation intensity ($\mathcal{C}\approx45$), whereas suicide and accident rates displayed higher and similar values  ($\mathcal{C}\approx95$). Over time, $\mathcal{C}$ for accidents decreased to a plateau of $\mathcal{C}\approx30$ after 2012, while for suicides, it slightly declined to a plateau of $\mathcal{C}\approx75$. In contrast, homicide rates exhibited a marked increase between 2005 and 2017, especially after 2012, rising to a plateau of $\mathcal{C}\approx220$ -- almost a five‐fold increase over a period slightly exceeding one decade.

In addition, we estimate the evolution of Moran's $I$~\cite{moran1950notes}, a widely used measure of spatial autocorrelation~\cite{getis2008history} corresponding to a spatially-weighted extension of the Pearson correlation (see \hyperref[sec:methods_moran]{Methods} for details). Figure~\ref{fig:8}B shows that the evolution of Moran's $I$ corroborates the findings from the correlation intensity $\mathcal{C}$. As in the previous analysis, Moran's $I$ decreased for suicide and accident rates, stabilizing at levels slightly higher for suicides than for accidents. Conversely, Moran's $I$ also increased significantly after 2004 for homicides, plateauing after 2017. For both $\mathcal{C}$ and Moran's $I$, we find no significant differences between the spatial behavior observed of accident rates and their disaggregated categories, although non-transport accidents displayed slightly lower values in years before 2010 (Supplementary Figure~S20).

\section*{Discussion and Conclusions}
Our analysis shows that the relationship between city size and external causes of death in Brazil has evolved substantially over the past three decades. Integrating urban scaling theory with spatial correlation analyses, we find that homicide, suicide, and accident fatalities each display distinct scaling regimes whose temporal dynamics reveal complex underlying processes in urban systems. For homicides, the persistence of superlinear scaling -- associated with pronounced agglomeration effects in enhanced social interactions in large urban areas -- has diminished in recent years. The scaling exponent declined from a high of $\beta=1.33$ in 2012 to approximately $\beta=1.12$ in 2022, a trend paralleled by a marked increase in spatial correlation of homicide rates. Suicide fatalities, traditionally characterized by sublinear scaling -- implying a protective effect associated with urban agglomeration -- have gradually become less sublinear over time, rising from $\beta=0.70$ in 1999 to $\beta=0.92$ in 2022. Accident-related deaths exhibit a persistent superlinear scaling, with an overall exponent of approximately $\beta=1.1$ over the study period. However, divergent behaviors emerge when accidents are disaggregated into transport and non-transport categories. Transport accidents consistently scale isometrically. Conversely, a superlinear regime emerged for non-transport accidents after 2010. Different from homicides, the trends observed for suicides and accidents were accompanied by weaker and stable spatial correlations.

Taken together, these findings suggest that distinct mechanisms underlie the observed changes in the external causes of death. For homicides, the decline in scaling exponents (a reduction in superlinearity) coincides with the period after 2012, when spatial correlations increased more rapidly. This finding indicates that homicides are becoming more spatially widespread, shifting from a concentration in large urban centers toward an increasing prevalence in smaller cities. Simultaneously, homicide rates in large cities -- despite typically being separated by considerable distances -- are becoming more spatially clustered over time, resulting in higher correlations at larger distances and suggesting the intensification of inter-city interactions underlying these violent deaths. Although our analysis is not causal, the observed spatiotemporal patterns align with the hypothesis that organized crime has consolidated into nationwide networks~\cite{stahlberg2022prison}. Such consolidation may reduce violence in ways that ultimately benefit criminal enterprises, particularly those involved in drug trafficking, and increased coordination among organized crime groups is one potential mechanism that could contribute to the observed rise in the spatial correlation of homicide rates. Moreover, several other, potentially interacting, mechanisms may contribute to explaining the homicide patterns. The diffusion of violence from large to smaller cities, likely facilitated by improved transport and communication infrastructures, can yield long-range and lead–lag dynamics in which changes in one city are followed by changes in nearby cities. Another contributing factor could be coordinated policy interventions or common shocks, such as shifts in policing, judicial practices, or social programmes, that can affect multiple cities simultaneously. Additionally, localized policies targeting crime prevention in large urban centers can inadvertently displace criminal activities to neighboring or more vulnerable regions. Finally, structural socioeconomic similarities across cities can create conducive environments for diffusion or displacement of violence, leading to the emergence of regionally distinct clusters of violence. Testing these mechanisms and disentangling their relative contribution, however, lies beyond the scope of this study, and we hope future research can address these issues. 

In contrast, for suicides, the steady increase in scaling exponents occurs alongside almost stable spatial correlations. Similarly, the changes in scaling exponents for non-transport accidents after 2010 do not appear to be associated with the small variations in spatial correlations. As with homicides, it remains challenging to firmly ascertain the underlying mechanisms driving these changes; however, the less correlated and stable spatial patterns suggest that the processes involved occurred within city boundaries. Possible explanations for the observed changes in suicides include a likely heightening of urban stressors in larger cities -- such as intensified economic competition, social isolation, and environmental pressure -- which may elevate suicide risk among residents of large cities. Additionally, shifts in cultural and social conditions~\cite{durkheim1897}, including evolving social networks and community structures~\cite{fowler2008dynamic, hill2010emotions}, could exacerbate feelings of anonymity and disconnection~\cite{durkheim1897} in densely populated areas, whereas in smaller cities, greater engagement with social networks might foster connectedness~\cite{melo2014statistical}. In the case of non-transport accidents, potential explanations include the expansion of informal employment in large urban areas -- with its attendant reduced adherence to safety standards and oversight -- periods of intensive construction driven by rapid urban expansion, and climatic variability and extreme weather events that may disproportionately affect densely populated areas. Moreover, the relatively stable and weaker spatial correlations observed for suicides and accidents further suggest that these changes are more closely linked to within-city dynamics than to inter-city interactions. 

Our regional analyses further reveal that, although inter-state variability exists in both the scaling exponents and the pace of change, state-specific exponents follow the country's overall behavior. In contrast, substantial heterogeneity is observed in scale-adjusted mortality indicators and their evolution across Brazilian states for all external causes of death. Cities in certain states consistently register fatalities above or below those predicted by the scaling laws and display distinct average scale-adjusted trends, including transitioning from negative to positive or near zero -- or vice versa -- over time. This variability across states underscores that local policies, cultural practices, socioeconomic conditions, and public health interventions can significantly impact the relative number of deaths in cities, while the robust and homogeneous evolution of scaling exponents suggests that changes in scaling laws are driven by more fundamental socioeconomic, infrastructural, and organizational processes at the national level.

Although our study leverages a comprehensive dataset and employs robust Bayesian hierarchical methods, it is not without limitations. While Zipf's Law was applied to mitigate potential issues arising from small urban settlements that may lack the complexity and agglomeration effects typical of urban areas, Brazilian cities are defined as administrative units that may not fully capture the functional boundaries of urban agglomerations. Moreover, even though our spatial correlation analysis elucidates patterns of inter-city dependence, we emphasize that they are not enough for establishing causal links between these patterns and specific socioeconomic or policy interventions underlying the changes in the scaling law remains challenging. The mechanisms we discuss are plausible -- though untested -- hypotheses and should be interpreted in light of these limitations. Despite these constraints, our findings indicate that the scaling relationships governing external causes of death are dynamic, reflecting both structural changes at the national level and regional disparities. The weakening of superlinear scaling for homicides and sublinear scaling for suicides, the emergence of superlinear scaling for non-transport-related accidents, and the intensified spatial clustering of violent deaths signal a fundamental shift in the urban fabric of Brazil. As cities continue to expand and evolve, our results underscore the imperative for dynamic, context-specific strategies to address the multifaceted challenges of urban violence, mental health, and public safety. Future research integrating refined urban delineations and additional socioeconomic variables will be critical for further unraveling these complex dynamics and informing effective interventions.

\section*{Methods}
\subsection*{Data}\label{sec:methods_data}
The data used in our investigation were sourced from the Brazilian Mortality Information System (\textit{Sistema de Informa\c{c}\~ao sobre Mortalidade} -- SIM)~\cite{mathers2005counting}, a nationwide database established by the Brazilian Ministry of Health in 1975 and currently managed under the Unified Health System (\textit{Sistema \'Unico de Sa\'ude} -- SUS)~\cite{castro2019brazil}. This platform consolidates mortality data from all Brazilian cities, serving as a critical resource for epidemiological research and health management. Data extraction from the Mortality Information System was conducted using the Python module \texttt{PySUS}~\cite{coelho2021pysus}, which provides tools for downloading and preprocessing datasets from Brazil's Unified Health System. Mortality records were then classified according to the underlying cause of death using the International Classification of Diseases, 10th Revision (ICD-10)\cite{who1992icd10}, yielding annual counts of deaths attributed to homicides (intentional deaths caused by another person; ICD-10 underlying cause codes: X85-Y09, Y87.1), suicides (intentional self-inflicted deaths; ICD-10 underlying cause codes: X60-X84, Y87.0), and accidents (unintentional fatal injuries; ICD-10 underlying cause codes: V01-X59, Y85-Y86) for each Brazilian city from 1996 and 2023. Unintentional deaths are further disaggregated into transport (ICD-10 underlying cause codes: V01-V99, Y85) and non-transport (ICD-10 underlying cause codes: W00-X59, Y86) accidents for additional analysis. Population data for the same 28-year period, along with spatial coordinates (latitude and longitude) for each city, were obtained from the Brazilian Institute of Geography and Statistics (\textit{Instituto Brasileiro de Geografia e Estat\'istica} -- IBGE)\cite{ibge}, the country's official statistical agency.

\subsection*{Bayesian hierarchical scaling model}\label{sec:methods_model}

To fully specify the hierarchical scaling model presented in Equation~\ref{eq:bayesian_scaling}, we adopt non-informative prior distributions to avoid introducing bias into the posterior estimates~\cite{gelman2006priordistributions}. Specifically, we use the following prior specifications: the residual variance $\varepsilon$ was assigned a uniform distribution, $\varepsilon \sim \mathcal{U}(0,10^2)$; the scaling parameters $\beta$ and $a$ were given normal distributions, namely $\beta \sim \mathcal{N}(0, 10^5)$ and $a \sim \mathcal{N}(0, 10^5)$; and their standard deviations $\sigma_\beta$ and $\sigma_a$ were modeled using inverse-gamma distributions, $\sigma_\beta \sim \text{Inv-}\Gamma(10^{-3}, 1)$ and $\sigma_a \sim \text{Inv-}\Gamma(10^{-3}, 1)$. Here, $\mathcal{U}(x_{\text{min}},x_{\text{max}})$ denotes a uniform distribution bounded between $x_{\text{min}}$ and $x_{\text{max}}$, and $\text{Inv-}\Gamma(c,d)$ represents an inverse-gamma distribution parameterized by shape $c$ and scale $d$. Bayesian inference was performed using the PyMC3 computational framework~\cite{pymc3}, which implements gradient-based Hamiltonian Monte Carlo (HMC) sampling with the No-U-Turn Sampler (NUTS) algorithm. To ensure reliable estimation and robust posterior sampling, we executed eight parallel Markov chains, each consisting of $10{,}000$ iterations, preceded by an initial burn-in phase of $5{,}000$ iterations, which were discarded. Convergence of the posterior distributions was evaluated using the Gelman-Rubin statistic ($\hat{R}$)~\cite{gelman1995bayesian}, confirming convergence with values close to $1$ across all model parameters and years. Posterior distributions, means, standard deviations, and 95\% credible intervals for the scaling exponents ($\beta$, $\beta_j$) and intercepts ($a$, $a_j$), were computed to quantify urban scaling relationships at both the national and state levels. 

\subsection*{Spatial autocorrelation function}\label{sec:methods_correlation}

We define the spatial correlation function as
\begin{equation}
C(r) = \frac{1}{M(r)} \sum_{i,j} \frac{(y_i - \bar{y})(y_j - \bar{y})}{\sigma^2} \, \chi\Big(|r_{ij}-r| < \frac{\Delta r}{2}\Big),
\end{equation}
where $y_i$ and $y_j$ denote the mortality rates in cities $i$ and $j$, respectively; $\bar{y}$ and $\sigma^2$ are the mean and variance of the mortality rates; and $r_{ij}$ is the distance between cities $i$ and $j$. The function $\chi$ is the indicator function defined by 
\begin{equation*}
\chi(\mathcal{E}) = \begin{cases}
1, & \text{if the condition $\mathcal{E}$ is true},\\
0, & \text{otherwise}.
\end{cases}
\end{equation*}
Thus, $\chi(|r_{ij}-r| < {\Delta r}/{2})=1$ precisely for city pairs whose separation falls inside the radial bin $[r- {\Delta r}/{2},r+ {\Delta r}/{2}]$ and 0 otherwise. In turn, the normalizing factor $M(r)=\sum_{i,j}\chi(\big|r_{ij}-r\big| < {\Delta r}/{2})$ is the total number of pairs in that bin. In our analyses, distance bins were selected based on the percentiles of the inter-city distance distribution to ensure that $C(r)$ is calculated from a comparable number of city pairs.

\subsection*{Moran's spatial autocorrelation index}\label{sec:methods_moran}
Moran's $I$~\cite{moran1950notes,getis2008history} spatial autocorrelation measures is defined as
\begin{equation}
    I = \frac{M}{W_0} \frac{\sum_i^M\sum_j^M w_{ij} (y_i-\bar{y})(y_j-\bar{y})}{\sum_i^N(y_i-\bar{y})^2 }\,,
\end{equation}
where $y_i$ and $y_j$ denote the mortality rates in cities $i$ and $j$, $\bar{y}$ represent the average mortality rates across all $M$ cities, $W_0=\sum_i^M\sum_j^M w_{ij}$ is a normalization constant, and $w_{ij}$ is the weight assigned to each city pair. Analogous to the Pearson correlation coefficient, Moran's $I$ typically ranges from $-1$ (indicating strong negative spatial autocorrelation) to $1$ (indicating strong positive spatial autocorrelation)~\cite{de1984extreme}, with values near zero suggesting spatial randomness~\cite{de1984extreme}. In our analysis, the spatial weight matrix $w_{ij}$ is specified as an inverse function of the square root of the intercity distance $r_{ij}$, $w_{ij}=1/\sqrt{r_{ij}}$, to assign a relatively high weight to distant cities; however, alternative choices, such as setting $w_{ij}=1$ for the first $k$ nearest neighboring cities and $w_{ij}=0$ otherwise, yield similar results (Supplementary Figure~S21).

\section*{Funding Declaration}
M.V.P. and H.V.R. acknowledge the support of the Coordena\c{c}\~ao de Aperfei\c{c}oamento de Pessoal de N\'ivel Superior (CAPES) and the Conselho Nacional de Desenvolvimento Cient\'ifico e Tecnol\'ogico (CNPq -- Grant 303533/2021-8). J.S.A. acknowledges funding from Fundação Cearense de Apoio ao Desenvolvimento Cient\'ifico e Tecnol\'ogico (FUNCAP, Award 06849573/2023), Conselho Nacional de Desenvolvimento Cient\'ifico e Tecnol\'ogico (CNPq, Award 303765/2017-8, Bolsa PQ), and Coordena\c{c}\~ao de Aperfei\c{c}oamento de Pessoal de N\'ivel Superior (CAPES, Award 88887.311932/2018-00, CAPES/PRINT). The funders had no role in study design, data collection and analysis, decision to publish, or preparation of the manuscript.\clearpage

\section*{Author contributions}
C.I.N.S, H.A.C., A.S.L., M.V.P., H.V.R., M.C.C., and J.S.A. designed research, performed research, analyzed data, and wrote the paper.

\section*{Data availability}
The data used during the current study are freely available from the Brazilian Mortality Information System (\textit{Sistema de Informa\c{c}\~ao sobre Mortalidade} -- SIM)~\cite{mathers2005counting} and Brazilian Institute of Geography and Statistics (\textit{Instituto Brasileiro de Geografia e Estat\'istica} -- IBGE)\cite{ibge}, as well as from the corresponding authors on reasonable request.

\bibliography{references}

\clearpage
\includepdf[pages=1-22,pagecommand={\thispagestyle{empty}}]{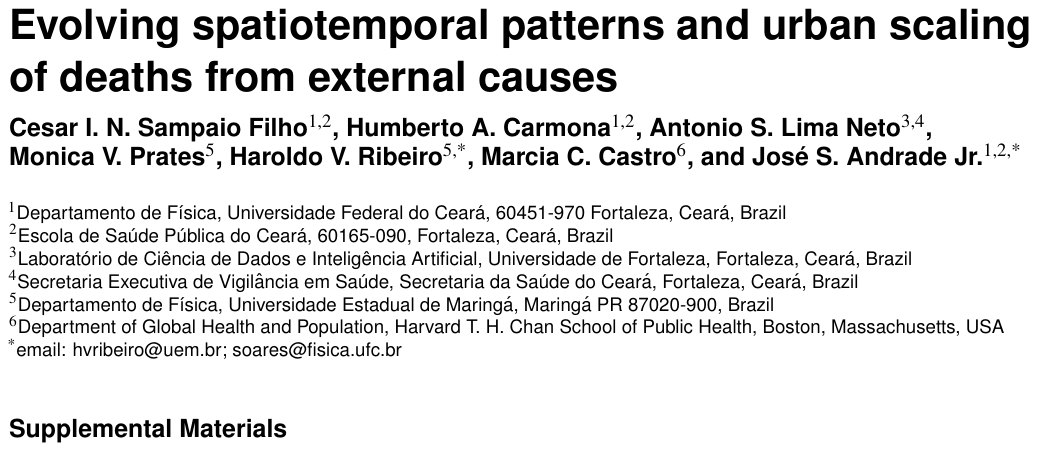}

\end{document}